# Electrowetting on dielectrics on lubricating fluid based slippery surfaces with negligible hysteresis

J. Barman[a], A. K. Nagarajan[b] and K. Khare[a*]

[a]*Department of Physics, Indian Institute of Technology Kanpur, Kanpur-208016, India*

[b]*Hindustan Unilever Research Centre, Banglore-560066, India*

*Email: kcharya@iitk.ac.in*

Low voltage electrowetting on dielectrics on substrates with thin layer of lubricating fluid to reduce contact angle hysteresis is reported here. On smooth and homogeneous solid surfaces, it is extremely difficult to reduce contact angle hysteresis (contact angle difference between advancing and receding drop volume cycle) and the electrowetting hysteresis (contact angle difference between advancing and receding voltage cycle) below 10°. On the other hand, electrowetting hysteresis on rough surfaces can be relatively large (>30°) therefore they are of no use for most of the fluidic devices. In the present report we demonstrate that using a thin layer of dielectric lubricating fluid on top of the solid dielectric surface results in drastic reduction in contact angle hysteresis as well as electrowetting hysteresis (< 2°) on smooth as well as rough surfaces. Subsequently fitting the Lippmann-Young electrowetting equation to the experimental electrowetting data reveal that the dielectric lubricating fluid layer is only responsible for smooth movement of the three phase contact line of the liquid drop and does not affect the effective specific capacitance of the system.

**Introduction:**

Electrowetting on dielectrics (EWOD) has evolved as one of the most efficient technique to manipulate interfacial tension as well as apparent contact angle and is also used in different applications like microfluidics or Lab-on-a-chip[1-5], variable focal length liquid lens[6-8], electrowetting display[9,10] to name a few. In EWOD, reversible modulation of apparent contact angle of a conductive liquid drop is obtained by applying voltage between a conductive substrate and the conducting liquid drop which are separated by a thin hydrophobic dielectric layer.[11,12] Quantitatively, variation of contact angle with voltage during electrowetting was first given by Lippmann and Young[11]

$$\cos\theta_V = \cos\theta_Y + \frac{CV^2}{2\gamma} \qquad (1)$$

where $\theta_V$ is the apparent contact angle at voltage $V$ during electrowetting, $\theta_Y$ is the Young's contact angle without voltage, $C = \epsilon_0\epsilon_r/d$ (where $\epsilon_0$ is permittivity of the vacuum, $\epsilon_r$ is the dielectric constant of material, $d$ thickness of the dielectric layer) is the specific capacitance due to the dielectric layer, $\gamma$ is the surface tension of conductive liquid and $V$ is the applied ac voltage. Below the contact angle saturation voltage, which depends on the material's properties, EWOD works very well thus widely used in many applications and devices. It has been demonstrated that using dielectrics of high dielectric constant and very low thickness results in more efficient electrowetting behavior.[13, 14] But still the biggest challenge in EWOD experiments is the electrowetting hysteresis, i.e. the difference between contact angles for increasing and decreasing voltage cycles after reaching back to 0V. As of now, it is extremely difficult to bring the electrowetting hysteresis below 10° in normal electrowetting experiments. In the last decade, several research groups demonstrated how to minimize the electrowetting hysteresis on smooth as well as rough surfaces using various approaches. Mugele *et al.* showed that the electrowetting hysteresis on a moderately rough surface could be decreased substantially using ac voltage whereas dc voltage has no significant effect on it.[15] They concluded that electrowetting experiments at low applied voltages show a large electrowetting hysteresis compared to large voltages. Berge *et al.* achieved very low electrowetting hysteresis (~ 2°) for oil drops under an immiscible conducting liquid (reverse electrowetting) on substrate with *rms* surface roughness of 100 nm.[16] However, no study has been done to decrease the electrowetting hysteresis in ambient (air) medium substantially on substrates with low/high roughness which can be very useful in many electrowetting based devices.

Aizenberg *et al.* recently demonstrated the slippery behaviour of a test liquid on lubricating fluid impregnated porous surfaces.[17-19] Due to the lubricant fluid coating on smooth and/or rough surfaces, the three phase contact line of a test liquid drop is free to move therefore can easily slip on these surfaces upon tilting by few degree. Due to smooth motion of the three phase contact line of a test liquid drop, the contact angle hysteresis on lubricating fluid impregnated smooth/rough surfaces is extremely low,[18] which inspired us to use this system for EWOD experiments to reduce the electrowetting hysteresis.

**Experimental setup:**

In the electrowetting on dielectric (EWOD) experiments, single side polished p-type <100> silicon substrates were used as conducting substrates. *rms* surface roughness on polished and rough sides of the substrates were found to be 1.2 nm and 183 nm respectively as measured by an Atomic Force Microscope ((PicoSPM, Non-contact mode, Molecular Imaging Corporation). Thin insulating layer (1 µm ± 0.15) of silicon dioxide ($SiO_2$) was grown thermally on both sides of silicon substrates and was used as a dielectric layer. EWOD experiments were carried out on both side of the silicon substrates with and without lubricating fluid. To start with large initial contact angle on silicon substrates during EWOD, self-assembled monolayer of Octadecyltrichlorosilane (OTS) molecules was grafted following standard procedure.[20] Mixture of water, glycerol and salt (NaCl) (17:80:3) was used as a test (conductive) liquid and 10 kHz frequency ac voltage was used for the EWOD experiments.[21] Contact angles of the test liquid drops as a function of applied ac voltages were measured from images obtained by a contact angle goniometer (OCA-35, DataPhysics, Germany) after 2 sec interval so that the test liquid had sufficient time to go to equilibrium. Silicone oil (viscosity ≈ 350 cSt) was used as a lubricating fluid and was spin coated on polished and rough silicon substrates to cast as thin liquid film. Thickness of the lubricating fluid layer was measured as approximately 3 µm by measuring the weight difference before and after the coating silicone oil on the substrates.

**Results and discussions:**

Figure 1(a) and (b) shows the plot of the apparent contact angle of the test liquid drop as a function of applied ac voltage for the 1$^{st}$ cycle of increasing and decreasing voltage on the smooth and rough substrates respectively. Upon increasing voltage, due to repulsion of like induced charges in the polarized dielectric $SiO_2$ layer results in decrease of the solid-liquid interfacial energy ($\gamma_{SL}$), hence decreases the apparent contact angle ($\theta_V$).

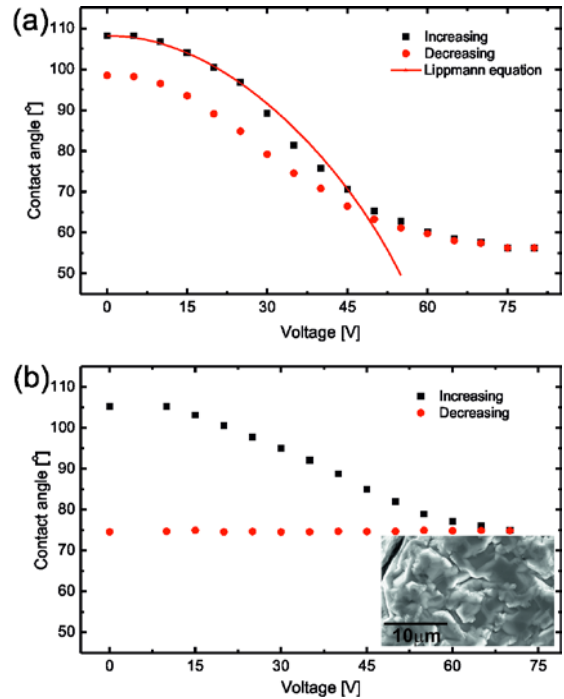

FIG.1. Plot of apparent contact angle ($\theta_V$) as a function of applied ac voltage for EWOD experiment on (a) smooth and (b) rough substrates. Red solid line in (a) represents Eq. 1. Inset in (b) shows SEM image of a rough silicon substrate.

On smooth substrates, the apparent contact angle decreases from 109° to 56° after applying 75 volts and becomes constant indicating contact angle saturation.[11, 22] Black solid curve in Fig. 1(a) represents Lippmann-Young equation (Eq. 1) and black and red dots represents experimental data points corresponding to increasing and decreasing voltage cycle. Equation 1 was fitted to the experimental data (excluding the saturation region) for increasing voltage cycle and the dielectric constant of dielectric layer was used as a fitting parameter which came out as 4.1 which is almost same as the bulk dielectric constant of pure silicon dioxide (3.9). Upon decreasing the voltage, the apparent contact angle increased and reached back to 99° at 0 V resulting the electrowetting hysteresis of 10°. Figure 1(b) shows the EWOD experiment on the rough substrates for increasing and decreasing voltage cycles. Due to very large roughness (*rms* value 183 nm) on rough substrates there are large number of pinning sites as clearly seen in the Scanning Electron Micrograph shown in Fig. 1(b) inset. That is why as the three phase contact line of the test liquid drop gets pinned upon changing the voltage and the net electrowetting behavior appears quite different compared to the smooth surface. On rough substrates, the apparent contact angle decreased from 109° to 75° upon applying 70 V and saturated after that. Due to random pinning of the three phase contact line upon increasing voltage, it was not possible to fit the Lippmann-Young equation for this case. Upon decreasing the voltage back to 0 V, the apparent contact angle did not increase at all due to complete pinning which resulted in 75° contact angle at 0 V. Therefore on rough surfaces the electrowetting hysteresis was about 35° which is a relatively large value. Therefore, rough surfaces can hardly be used for any practical fluidic device applications where a liquid has to move on a solid surface say via electrowetting.

Primary reason of the contact line pinning is topographical and chemical heterogeneity on solid surfaces. If these topographical and/or chemical heterogeneities is covered by a coating which provides smooth movement of the three phase contact line of a drop, contact angle hysteresis would decrease significantly. It has already been demonstrated that coating a surface by a thin lubricating fluid makes a test drop slip upon tilting the substrate. Hence it is expected that lubricating fluid coated substrates would also reduce the contact angle hysteresis due to smooth movement of the three phase contact line on the lubricating film. Therefore we coated the smooth as well as rough substrates with thin film of silicone oil as lubricant. It was observed that if silicone oil is directly coated on silicon substrates (which are primarily hydrophilic), test liquid drops show very poor stability as they quickly sink inside the lubricant film and become immobile. It has recently been shown that liquid drop on top of a non-miscible fluid layer depicts Neumann's contact angle rather than Young's contact angle.[23] Since the silicon substrates are primarily hydrophilic, the test drop prefers to wet the substrate even if they are separated by the lubricating film. Therefore the substrate has to be hydrophobic to avoid the test drop sinking in the lubricating film. So thin lubricating films of silicone oil were coated on hydrophobic (with self-assembled monolayer of OTS molecules) silicon substrates. These lubricating fluid coated hydrophobic silicon substrates (smooth and rough) were used for all subsequent experiments. Figure 2(a) shows a contact angle hysteresis plot on a lubricating fluid coated rough Si substrate by increasing and decreasing volume of the test fluid. This clearly indicates that the contact angle hysteresis on a silicone oil coated substrates is around 2°. This is due to the fact that the lubricating surface provides smooth motion of the three phase contact line of the test liquid drops upon increasing and decreasing drop volume therefore showing very low contact angle hysteresis. Subsequently electrowetting experiments were performed on silicone oil coated smooth and rough samples. Figure 2(b) shows EWOD response of a test liquid drop on the lubricating fluid coated rough substrate. Qualitative the EWOD behavior appears very similar to the one on solid hydrophobic substrates (c.f. Fig 1(a)), quantitatively it is very different. On lubricating fluid coated substrates there are mainly three transition regions. The apparent contact angle upon electrowetting does not change with voltage up to 25V, clearly showing a threshold voltage ($V_{th}$) behaviour which is also reported by Bormashnako et al. for electrowetting response on silicon oil impregnated polycarbonate structured surfaces.[13] Beyond the threshold voltage $V_{th}$, the apparent contact angle decreases with increasing voltage following the conventional Lippmann-Young behaviour. Beyond 60V the apparent contact angle saturates and not depends upon the applied voltage depicting contact angle saturation. The contact angle saturation phenomenon on lubricating fluid coated solid surfaces is similar to the one seen on normal hydrophobic solid substrates and have been explained by various theoretical models.[11,22] Upon decreasing the voltage (receding voltage cycle), the apparent contact angle increases following the same Lippmann-Young behaviour up to the threshold voltage $V_{th}$. Decreasing the voltage below the threshold voltage does not increase the contact angle any further. This results in electrowetting hysteresis around 2° which is almost negligible compared to the total change of contact angles in the increasing and decreasing voltage cycle of the electrowetting experiment. Black solid line in Fig. 2(b) indicates the Lippmann-Young equation (Eq.1) which was fitted numerically to the experimental data (advancing voltage only) with specific capacitance of the system as the fitting parameter (contact angles below the threshold voltage were neglected in the fitting routine). In this case also, since the lubricating fluid covers all the chemical and/or topographical inhomogeneities, the three phase contact line of the test fluid can move freely while increasing and decreasing the applied voltage. Electrowetting experiments performed on smooth substrates, coated with the same lubricating fluid, gave the same results. This is also expected as the smooth Si substrates don't have any topographical pinning sites and the lubricating fluid smoothens the motion of the three phase contact line of drops of the test liquid during electrowetting. The only difference of electrowetting on pure solid and lubricating fluid coated substrates is the onset of the response of apparent contact angle with applied voltage (threshold behaviour). The origin of the threshold behaviour on lubricating fluid surface (compared to the pure solid surface) is due to the fact that only $SiO_2$ is the dielectric layer in pure solid surface where as the combined layers of $SiO_2$ and silicone oil forms the net dielectric layer for lubricating fluid surfaces. Therefore the specific capacitance in case of lubricating fluid surfaces is the linear combination of the one due to silicone oil and the one due to $SiO_2$. Since the dielectric silicone oil has lower polarizability compared to the dielectric $SiO_2$, it requires larger voltage for the polarization at the liquid-liquid interface. Therefore the net dielectric response of the lubricating fluid surfaces shows the threshold characteristic which is reflected as threshold behaviour in the apparent contact angle.

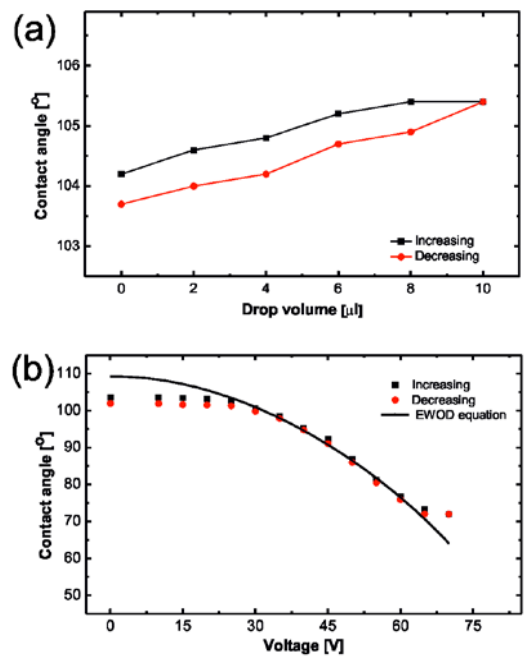

FIG.2. (a) Contact angle hysteresis on a lubricant coated rough Si substrate and (b) Electrowetting hysteresis curve on lubricant coated rough Si substrate.

Lubricating fluid coated solid surfaces (both rough and smooth) can be modelled as two series capacitors, one each for silicone oil and $SiO_2$, with $C_1$ and $C_2$ as specific capacitance respectively. So the net specific capacitance of the system can be calculated as

$$C = \frac{C_1 C_2}{C_1 + C_2} = \frac{\epsilon_0 \epsilon_1 \epsilon_2}{(\epsilon_1 d_2 + \epsilon_2 d_1)} \qquad (2)$$

where $\epsilon_1$, $d_1$ and $\epsilon_2$, $d_2$ are the dielectric constant and thickness of the silicone oil and SiO$_2$ respectively. From fitting the Lippmann-Young equation in the electrowetting curve of Fig. 2(b), the total specific capacitance was calculated as $C = (2.19 \pm 0.02) \times 10^{-5} F/m^2$. However the specific capacitance calculated using Eq. 2 resulted as $C = 5.61 \times 10^{-6} F/m^2$ which is almost four times smaller than the value obtained from the fitting. Assuming only SiO$_2$ layer responsible for electrowetting phenomenon, the specific capacitance results in the value $C = 3.45 \times 10^{-5} F/m^2$ which is very close to the fitting value. Therefore the net capacitance of the lubricating fluid based substrates is dominated by the silicon dioxide dielectric layer and the role of the lubricating fluid film is only to provide the chemically and topographically smooth surface for the free motion of the three phase contact line of a test liquid drop to decrease the electrowetting hysteresis. So it will be very effective during electrowetting experiments to have a thin layer of dielectric lubricant coating to reduce the electrowetting hysteresis to negligible value without affecting the total electrowetting behavior.

**Conclusions:**

In summary, we have demonstrated how to decrease the contact angle hysteresis as well as the electrowetting hysteresis on a smooth and rough surface with application of a thin layer of a dielectric lubricating film. Even for smooth solid surfaces, there is always finite electrowetting hysteresis. On rough or chemically heterogeneous solid surfaces the electrowetting hysteresis can be extremely large. With the help of a dielectric lubricating fluid, which has to be immiscible to the test liquid, the contact angle hysteresis as well as the electrowetting hysteresis can be brought down to negligible value. This is due to the fact that the lubricant fluid covers all chemical and topographical inhomogeneities and provides smooth movement of the three phase contact line of a test liquid drop thus reducing the hysteresis. Electrowetting behaviour on a pure solid surface is quantitatively explained by Lippmann-Young equation which depends upon the thickness and dielectric constant of the dielectric layer as well the surface tension of the test liquid. Solid substrates, coated with lubricant fluid, behaves like a two capacitors connected in series during electrowetting experiments in which the capacitance due to

$SiO_2$ dominates over the capacitance due to lubricant. Therefore the net specific capacitance of the system is primarily governed by the dielectric silicon dioxide layer as confirmed by fitting the Lippmann-Young equation to the electrowetting experiments and role of the lubricant is primarily to provide smooth movement of the three phase contact line of the test liquid drops.

**Acknowledgements:**

This research work was supported by Hindustan Unilever Limited, India and DST, New Delhi through its Unit of Excellence on Soft Nanofabrication at IIT Kanpur.